\documentclass[a4paper,fleqn,twocolumn,usenatbib]{mnras}

\usepackage{mathptmx}
\usepackage{ulem}
\usepackage[T1]{fontenc}
\usepackage{ae,aecompl}

\usepackage{graphicx}	
\usepackage{amsmath}	
\usepackage{amssymb}	
\usepackage{indentfirst}

\def\nn{\nonumber}

\def\be{\begin{equation}}
\def\ee{\end{equation}}
\def\beq{\begin{eqnarray}}
\def\eeq{\end{eqnarray}}
\def\nn{{\nonumber}}

\title[SMBH GALAXY CO-EVOLUTION]{Co-evolution of supermassive black holes with galaxies from semi-analytic model: stochastic gravitational wave background and black hole clustering}

\author[Q. Yang et al.]{
Qing Yang,$^{1}$
Bin Hu,$^{1}$\thanks{E-mail: bhu@bnu.edu.cn}
Xiao-Dong Li,$^{2}$\\
$^{1}$Department of Astronomy, Beijing Normal University, Beijing, 100875, China\\
$^{2}$School of Physics and Astronomy, Sun Yat-Sen University, Guangzhou 510297, P. R. China
}

\date{Accepted XXX. Received YYY; in original form ZZZ}

\pubyear{2018}

\begin{document}
\label{firstpage}
\pagerange{\pageref{firstpage}--\pageref{lastpage}}
\maketitle

\begin{abstract}
We study the co-evolution of supermassive black holes (SMBHs) with galaxies by means of semi-analytic model (SAM) of galaxy formation based on sub-halo merger trees built from Millennium and Millennium-II simulation. We utilize the simulation results from Guo 2013 and Henriques 2015 to study two aspects of the co-evolution, \emph{i.e.} the stochastic gravitational wave (GW) background generated by SMBH merger and the SMBH/galaxy clustering. The characteristic strain amplitude of GW background predicted by Guo 2013 and Henriques 2015 models are $A_{yr^{-1}}=5.00\times10^{-16}$ and $A_{yr^{-1}}=9.42\times10^{-17}$, respectively. We find the GW amplitude is very sensitive to the galaxy merger rate.
The difference in the galaxy merger rate between Guo 2013 and Henriques 2015, results in a factor $5$ deviation in the GW strain amplitude.
For clusterings, we calculate the spatially isotropic two point auto- and cross-correlation functions (2PCFs) for both SMBHs and galaxies by using the mock catalogs generated from Guo 2013 model. We find that all 2PCFs have positive dependence on both SMBH and galaxy mass. And there exist a significant time evolution in 2PCFs, namely, the clustering effect is enhanced at lower redshifts. Interestingly, this result is not reported in the active galactic nuclei samples in SDSS. Our analysis also shows that, roughly, SMBHs and galaxies, with galaxy mass $10^2\sim10^3$ larger than SMBH mass, have similar pattern of clustering, which is a reflection of the co-evolution of SMBH and galaxy.
Finally, we calculate the first ten multiples of the angular power spectrum of the energy density of GW background. We find the amplitude of angular power spectrum of the first ten multiples is about $10\%$ to $60\%$ of the monopole component in the whole frequency range.
\end{abstract}

\begin{keywords}
galaxies: formation, (galaxies:) quasars: supermassive black holes
\end{keywords}



\section{Introduction}
Observational evidence shows that supermassive black holes (SMBHs) are located in the center of nearly all massive galaxies \citep{Soltan:1982vf,Kormendy:1995er,Magorrian:1997hw}. Although the evolution mechanism of SMBHs is not very well known yet, observational evidence shows that there are strong correlations between mass of SMBHs and observational properties of their host galaxies, such as velocity dispersion, star formation rate and bulge stellar mass \citep{Madau:1996aw,Boyle:1997sm,Magorrian:1997hw,Ferrarese:2000se,Ueda:2003yx,Zheng:2009ac}. It is also expected that when galaxies merge, the SMBHs inside them should form SMBH binaries, emit gravitational waves during inspiral and merge eventually (we refer the readers to the latest review \cite{Sesana:2014wta}). Gravitational torques induced by galaxy-galaxy mergers drive inflows of cold gas toward the center of galaxies, triggering the
central starbursts and also accretion on to SMBHs \citep{Hernquist:1989ew,Barnes:1991zz,Barnes:1996qt,Mihos:1994wj,Mihos:1995ri,DiMatteo:2005ttp}. Galaxy/SMBH merger has been proposed to be a way by which central active galactic nuclei (AGN) could be triggered and SMBHs could grow \citep{Hopkins:2007hc,Sanders:1988rz,Treister:2012ag}.

Gravitational waves (GWs) from inspiralling SMBHs are expected to form a GW background at frequency range of $10^{-9}\sim 10^{-6}\mathrm{Hz}$. The detection of this GW background would have fundamental and far-reaching importance in cosmology and galaxy evolution not accessible by any other means. Precision timing of an array of millisecond pulsars (PTA) is a unique way to detect low frequency GW signal \citep{Sazhin:1978gy,Detweiler:1979wn,Blandford:1984hf,Foster:1990sl}; \citep{Blandford:1984hf}. Recently, European Pulsar Timing Array (EPTA) \citep{Ferdman:2010xq}, Parkes Pulsar Timing Array (PPTA) \citep{Manchester:2012xd} and North American Nanohertz Observatory for Gravitational Waves (NANOGrav) \citep{Jenet:2009xf}, joining together in the International Pulsar Timing Array (IPTA) \citep{Hobbs:2009yy}, are constantly improving their sensitivity in this frequency range, thus provide an important opportunity to get the very first low-frequency GW background detection. As PTA are promoting their upper limits on the GW background from SMBH mergers, several works have also reported their predictions on this GW background based on phenomenological models or simulations \citep{Jaffe:2002rt,Sesana:2012ak,Sesana:2016yky,Kelley:2016gse}. The predicted characteristic amplitude ($A_{yr^{-1}}$) is roughly in the range of $1\times10^{-16}$ to $5\times10^{-15}$.

The predictions of the characteristic amplitude focus on the isotropic property of the GW background. An isotropic GW background signal is an ideal case when the sources for the GW background has an infinite population, and is expected to produce the famous Hellings and Downs curve in the PTA observation \citep{Hellings:1983fr}. In realisty, the number of the SMBH binary is always finite, and the GW background signal they generated must has anisotropic components besides the dominated isotropic one. The anisotropic effect on PTA experiments and extension of the Hellings and Downs curve method to analyse anisotropies in the GW background has been developed in \cite{Cornish:2013aba}, \cite{Mingarelli:2013dsa}, \cite{Taylor:2013esa} and \cite{Cornish:2014rva}. At the same time, a first constraint on the anisotropy has been obtained with European Pulsar Timing Array data \citep{Taylor:2015udp}.

On the other hand, the galaxy/SMBH clustering may also provide a way to study the SMBH growth and its co-evolution with galaxy. Recent large-scale surveys, such as the Sloan Digital Sky Survey (SDSS), provide observational sample over hundreds of thousands AGNs \citep{Schneider:2010}. The auto-correlation of AGN and cross-correlation between AGNs and galaxies are studied with large samples \citep{Shen:2006ti,Shen:2008ez,Ross:2009sn,Coil:2009bi,Mountrichas:2008jf,Donoso:2009wd,Krumpe:2011ra,Komiya:2013vja,Shirasaki:2015lu}. The resulted correlations showed a positive dependence on BH mass and radio loudness, while no clear dependence was found on redshift or colour.

In this paper, we are going to utilize the semi-analytic galaxy formation model (SAM) based on sub-halo merger trees built from Millennium simulation \citep{Springel:2005nw}. We will focus on the Munich model, namely, Guo 2013 \citep{Guo:2010ap,Guo:2012fy} and Henriques 2015 \citep{Henriques:2014sga}, to make predictions on the rates at which SMBH form binaries and evolve to coalescence, the distribution of SMBH merger event, as well as the resulted characteristic strain amplitude of GW background and its anisotropic properties. We will also investigate the clustering property of both SMBHs and galaxies, as well as the cross-correlation between them with the mock catalogs generated from Guo 2013 \citep{Guo:2012fy}. The dependence of resulted correlation on redshift and BH/galaxy mass will be shown.

The rest of the paper is organized as follows. In section \ref{sec:gw}, we will first give the formula that we will use for the GW background, then briefly introduce the semi-analytical galaxy formation model, and compare the merger event distribution and merger rate derived from Guo 2013 and Henriques 2015. We will give our predictions on the characteristic strain amplitude derived from these two galaxy formation models, and compare them with previous results and PTA upper limits.
In section \ref{sec:correlation}, we will investigate clustering properties of both SMBHs and galaxies, and also the dependence of clustering amplitude on redshift and BH/galaxy mass.
Motivated by the results of the clustering property derived in the last section, In section \ref{sec:anisotropy}, we will study the anisotropy property of the GW background besides the monopole component by calculating the angular power spectrum of the energy density function.
Section \ref{sec:conclusion} is devoted to summaries and discussions.

\section{stochastic gravitational wave background}
\label{sec:gw}

\subsection{Gravitational wave strain}
In this subsection we review the formulae for GW background generated by a superposition of GW signal from SMBH binary sources \citep{Phinney:2001di,Sesana:2014wta}.
Consider the inspiral phase of SMBH binaries, without making any restrictive assumption about their kinetics, such as their semi-major axis and eccentricity evolution, we can write the characteristic strain spectrum $h_c^2$ of the background GW signal generated by the overall population as:
\beq\label{hcfirst}
h^2_c(f)&=&\int_0^{\infty}dz\int_0^{\infty}dM_1\int_0^1dq\frac{d^4N}{dzdM_1dqdt_r}\frac{dt_r}{d\mathrm{ln}f_{\mathrm{K},r}}\nn\\
&\times &h^2(f_{\mathrm{K},r})\sum_{n=1}^{\infty}\frac{g[n,e(f_{\mathrm{K},r})]}{(n/2)^2}\delta\left[f-\frac{nf_{\mathrm{K},r}}{1+z}\right]\;,
\eeq
where we have assumed the progenitor masses are $M_1$ and $M_2$ with $M_1>M_2$. Then $d^4N/dzdM_1dqdt_r$ is the differential cosmological coalescence rate of SMBH binaries per unit redshift ($z$), primary mass ($M_1$), mass ratio ($q=M_2/M_1<1$) and merge time ($t_r$). $dt_r/d\mathrm{ln}f_{\mathrm{K},r}$ is the time spent by the binary at each logarithmic frequency interval, where $f_{\mathrm{K},r}$ is the frequency of Keplerian motion measured in the rest frame of the binary. Together with $dt_r/d\mathrm{ln}f_{\mathrm{K},r}$, $d^4N/dzdM_1dqdt_r$ give the instantaneous population of orbiting binaries in a given logarithmic Keplerian frequency interval per unit redshift, primary mass and mass ratio. $h(f_{\mathrm{K},r})$ denotes the GW strain emitted by a circular binary at a Keplerian rest frame frequency $f_{\mathrm{K},r}$. The averaged strain over source orientations reads (see Thorne 1987 ``in Three Hundred Years of Gravitation'', ed. S. W. Hawking, W. Israel, Cambridge University Press)
\beq\label{hcircular}
h(f_{\mathrm{K},r})=\sqrt{\frac{32}{5}}\frac{\mathcal{M}^{5/3}}{D}(2\pi f_{\mathrm{K},r})^{2/3}\;,
\eeq
where $\mathcal{M}$ is the chirp mass, which is related to the progenitor masses by $\mathcal{M}=(M_1M_2)^{3/5}/(M_1+M_2)^{1/5}$, and $D$ is the luminosity distance to the source.
Equation \eqref{hcfirst} states that the background GW signal is a composition of the GW signal from each SMBH binary sources.
Note that having averaged over all the radiation orientation of the source, eq.~\eqref{hcircular} can be considered equivalently as an isotropic monopole radiation. The function $g(n,e)$ accounts for the fact that the binary radiates GW in the whole spectrum of harmonics $f_{r,n}=nf_{\mathrm{K},r}(n=1,2,...)$. In the circular case that we will consider throughout this paper, $g(n,e)=\delta_{n2}$, i.e. $g(n,e)=1$ for $n=2$, while is zero in other cases. The time duration spent on the logarithmic interval, $dt/d\mathrm{ln}f$, is given by the standard quadrupole formula as below \citep{Peters:1964zz}
\beq\label{dtdf}
dt/d\mathrm{ln}f=\frac{5}{64\pi^{8/3}}\mathcal{M}^{-5/3}f_r^{-8/3}\;.
\eeq
Besides these, we also have
\beq\label{dn}
\frac{d^2n}{dzd\mathcal{M}}=\frac{d^3N}{dzd\mathcal{M}d\mathrm{ln}f_r}\frac{d\mathrm{ln}f}{dt_r}\frac{dt_r}{dz}\frac{dz}{dV_c}\;,
\eeq
where $n$ is the comoving number density of coalescence, and $dV_c$ is the comoving volume shell lying between $z$ and $z+dz$. Plugging eq. \eqref{hcircular},\eqref{dtdf} and \eqref{dn} into \eqref{hcfirst}, we get the following background GW spectrum for the circular and quadrupole radiation
\beq\label{hc}
h_c^2(f)=\frac{4f^{-4/3}}{3\pi^{1/3}}\int\int dzd\mathcal{M}\frac{d^2n}{dzd\mathcal{M}}\frac{1}{(1+z)^{1/3}}\mathcal{M}^{5/3}\;.
\eeq
We see that in this case, $h_c\propto f^{-2/3}$, it is therefore customary to write the characteristic strain amplitude in the form $h_c=A(f/f_0)^{-2/3}$, where $A$ is the amplitude of the signal at the reference frequency $f_0$. Observational limits on the GW background are usually given in terms of $A$. Hereafter we denote $A$ with $f_0=1yr^{-1}$ as $A_{yr^{-1}}$.

\subsection{Semi-analytic model of galaxy formation}
In this subsection we'll first briefly introduce the general picture of semi-analytic model. And then, we will go into some details of the black hole self-regulated growth, in particular, the `quasar' and `radio' modes. Finally, we will compare the differences between the simulated results of Guo 2013 and Henriques 2015.

The semi-analytic model (SAM) of galaxy formation treats the baryonic evolution by post-processing cosmological N-body simulations in a way, that makes it possible to explore a wide model and parameter space in a reasonable amount of time. In this work, we utilize the Munich model/\texttt{L-Galaxies} code\footnote{\url{http://galformod.mpa-garching.mpg.de/public/LGalaxies/}}, which is based on the sub-halo merger trees built from the Millennium \citep{Springel:2005nw}/Millennium-II simulations \citep{BoylanKolchin:2009nc} (MS/MS-II), and applied to WMAP \citep{Guo:2010ap,Guo:2012fy}, Planck cosmology \citep{Henriques:2014sga} .
This model is developed based on a series of seminal works \citep{Springel:2005nw,Croton:2005fe,DeLucia:2006szx}. For readers who are interested in more details, we recommend to review papers \citep{Baugh:2006pf,Benson:2010ei,Benson:2010de}.
In the following paragraphs, due to the restriction of this topic, we very briefly summarize the general model and highlight the black hole self-regulated growth and relevant feedback mechanisms.

It is commonly believed that, galaxies form at the centers of dark matter halos. They gain stars by formation from interstellar medium (ISM) and by accretion of satellite galaxies. Galactic disc is formed from the materials in ISM. And those materials are replenished
both by diffuse infall from the surroundings and by gas from accreted satellite galaxies. There are two main channels for the diffuse infall. One is the direct infall of cold flow from intergalactic medium (IGM), the other is through cooling of the surrounded hot halos. Evolution of each galaxy is driven by the overall baryonic astrophysical complex network rather than a single process. This network includes not only the interactions among the processes mentioned above, but also the interactions of these processes with flows driven by SNe and by active galactic nuclei (AGN).

Due to the complexity of this system, our current understanding of most of these baryonic processes is mainly inspired by the simplified numerical simulations and by the phenomenology from observations. SAM may offer the best means to constrain them empirically using observational data.
The baryonic content of galaxies contains five components:  stellar bulge,  stellar disc, gas disc, hot gas halo as well as ejecta reservoir.
These components exchange materials through a variety of processes and gain mass via accreting IGM. The model parameters are estimated by using the observed abundance, structure and clustering of low-redshift galaxies as a function of stellar mass, luminosity and colour.

After reviewing the general picture, now we turn to the black hole self-regulated growth and feedback mechanisms. Following \cite{Croton:2005fe}, we can separate black hole growth into `quasar' mode and `radio' mode. The detailed recipes vary among different versions of the codes, here we take \cite{Guo:2010ap} as an example.

The `quasar' mode describes the black hole growth during gas-rich mergers.
During this process, the major black hole grows both by absorbing the minor and by accreting cold gas.
Hence, the final black hole mass can be expressed as
\begin{eqnarray}
M_{{\rm bh},f} & = & M_{{\rm bh, maj}}+M_{{\rm bh,min}}+\Delta M_{{\rm bh},Q}\;,\\
\label{eq:bhQ}\Delta M_{{\rm bh},Q} & = & \frac{f_{\rm bh}(M_{\rm min}/M_{\rm maj})M_{\rm cold}}{1+280~{\rm km}~{\rm s}^{-1}/V_{vir}}\;,
\end{eqnarray}
where $M_{{\rm bh, maj}}$, $M_{{\rm bh,min}}$, $M_{\rm cold}$, $V_{vir}$, $M_{\rm maj}$ and $M_{\rm min}$ are the black hole mass in the major and minor progenitors, the total cold gas in the two progenitors, virial velocity, the total baryon masses of the major and minor progenitors, respectively. $f_{\rm bh}$ is a free parameter, which is fixed to $0.03$ in order to reproduce the observed local $M_{\rm bh}-M_{\rm bulge}$ relation \citep{Croton:2005fe}.
Both major mergers and gas rich minor mergers contribute significantly to this channel. The feedback in the `quasar' mode is not explicitly written down in SAM. In some sense, we can treat the starburst as some indirect form of feedback in `quasar' mode.

`Radio' mode growth is through hot gas accretion on to central black holes. The rate in this mode is modelled as \citep{Croton:2005fe}

\begin{equation}
\label{eq:Mbhdot}\dot{M}_{{\rm bh}} = \kappa\left(\frac{f_{{\rm hot}}}{0.1}\right) \left(\frac{V_{vir}}{200~{\rm km}~{\rm s}^{-1}}\right)^3 \left(\frac{M_{{\rm bh}}}{10^8 h^{-1}M_{\odot}}\right)M_{\odot}~{\rm yr}^{-1}
\end{equation}
where $f_{\rm hot}$ is the ratio of hot gas mass to dark matter mass for the main subhalo case, and the ratio within some strip scales for a type-1 galaxy in a satellite subhalo case.
The parameter $\kappa$ sets the efficiency of hot gas accretion.
This hot gas accretion deposits energy in relativistic jets with $10\%$ efficiency. And this energy is transformed into heat in the atmosphere.
In \cite{Guo:2010ap}, the energy input rate is assumed as
\begin{eqnarray}
\dot{E}_{{\rm radio}} & = & 0.1\dot{M}_{{\rm bh}}c^2\;.
\label{eq:feedback}
\end{eqnarray}
Thus, the effective mass cooling rate is
\begin{eqnarray}
\dot{M}_{{\rm cool,eff}} & = & {\rm max}\left[\dot{M}_{{\rm cool}}-\frac{2\dot{E}_{{\rm radio}}}{V_{200c}^2},0\right]\;.
\label{eq:cooling}
\end{eqnarray}
It is the cooling onto the disk, the fuel of the star formation.  Eq. (\ref{eq:Mbhdot}) describes the growth/accretion of BH, and Eq. (\ref{eq:feedback}) describes the energy of AGN feedback in reheating the gas. Here reheating gas means to heat up the gas to virial temperature, that is why there is $V_{vir}$ appearing in Eq. (\ref{eq:Mbhdot}). Basically, $\dot{M}_{{\rm cool,eff}}$ describes the amount of gas that could have been cooled onto the disk if there were no AGN feedback. In the case of AGN feedback,  $2\dot E_{{\rm radio}}/V^2_{200c}$ amount of the cooling gas are reheated, so that the cooling rate is reduced according to Eq. (\ref{eq:cooling}).

In \cite{Henriques:2014sga} version, eq. (\ref{eq:bhQ}) and (\ref{eq:Mbhdot}) are replaced with
\begin{eqnarray}
\Delta M_{{\rm bh},Q} & = & \frac{f_{\rm bh}(M_{\rm min}/M_{\rm maj})M_{\rm cold}}{1+(V_{\rm bh}/V_{200c})^2}\;,\\
\dot{M}_{{\rm bh}} & = & k_{{\rm AGN}}\left(\frac{M_{{\rm hot}}}{10^{11}M_{\odot}}\right)\left(\frac{M_{{\rm bh}}}{10^8M_{\odot}}\right)\;,
\end{eqnarray}
and other equations keep the same.

After running SAM code, we get the whole galaxy merge history through the simulated time range. In the following, we will compare the simulation result of Guo 2013 and Henriques 2015 by mass-redshift distribution of SMBH merger events and galaxy merger rate. The SMBH merger history is related to the galaxy merger history assuming the two SMBHs will merge as soon as their host galaxies merged. Then both SMBH merger mass distribution and merger rate can be extracted from the simulated SMBH merger history.
\begin{figure}
\centering
\includegraphics[width=0.5\textwidth]{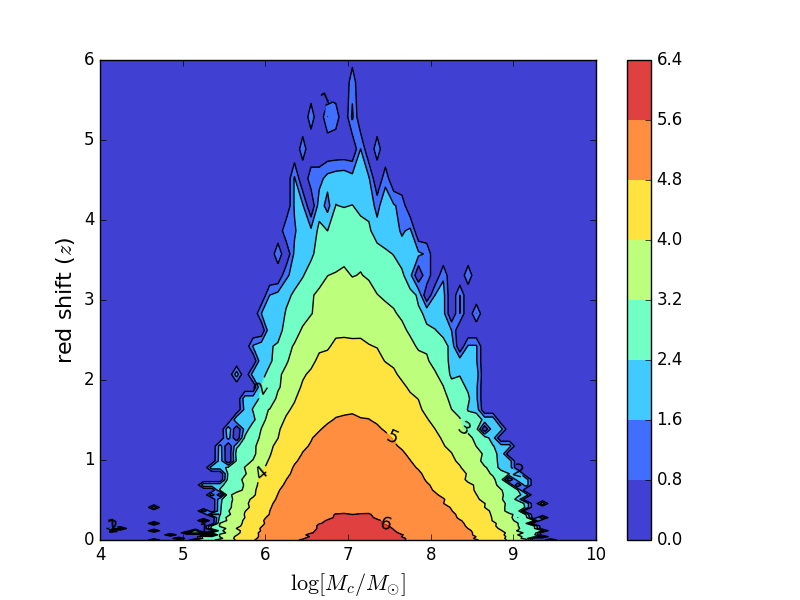}
\centering
\includegraphics[width=0.5\textwidth]{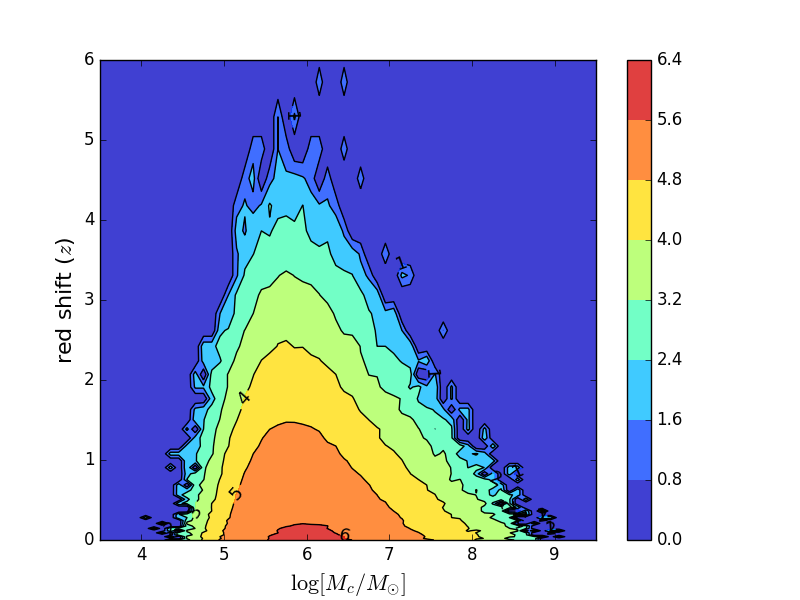}
\caption{\small The 2D contour plot the distribution of $d\log N(z,\mathcal{M})/(dzd\log (\mathcal{M}/M_{\odot}))$, namely, the logarithmic number of SMBH merger events from the whole box volume $V=[500\mathrm{Mpc}/h]^3$ per unit redshift and logarithmic mass-interval. Top panel: result from Guo 2013. Bottom panel: result from Henriques 2015.}\label{nummerger}
\end{figure}
In Fig.~\ref{nummerger} the logarithmic number of SMBH merger events from the whole box volume $V=[500\mathrm{Mpc}/h]^3$ per unit redshift and logarithmic mass-interval, i.e. $d\log N(z,\mathcal{M})/(d zd\log (\mathcal{M}/M_{\odot}))$, on redshift and logarithmic chirp mass plane are shown. 
We see that in both cases, SMBHs merge, on average, much more frequently in the low redshift region than in the high one. Its maximum locates around $z=0$, $\log\left(\mathcal{M}/M_{\odot}\right)=7$ (Guo 2013) and $\log\left(\mathcal{M}/M_{\odot}\right)=6$ (Henriques 2015), respectively. Then the differential event number gradually decrease to less than $10^{0.8}$ in the boundary of the plane, where either redshift is high or chirp mass is away from the maximum value. Furthermore, Fig.~\ref{nummerger} shows that SMBHs merger is more frequent in Guo 2013 than in Henriques 2015. The area of the maximum, i.e., the area where the differential number of merger events reaches about $10^6$, is larger in Guo 2013 than in Henriques 2015. Finally, there are more massive binaries, i.e. binaries with chirp mass larger than $10^9 M_{\odot}$ in Guo 2013 than in Henriques 2015, and the mass of the progenitor SMBHs is, on average, smaller in Henriques 2015. To sum up, systematically, the SMBHs merger is less massive and less frequent in Henriques 2015 than in Guo 2013.

\begin{figure}
  \centering
  \includegraphics[width=0.5\textwidth]{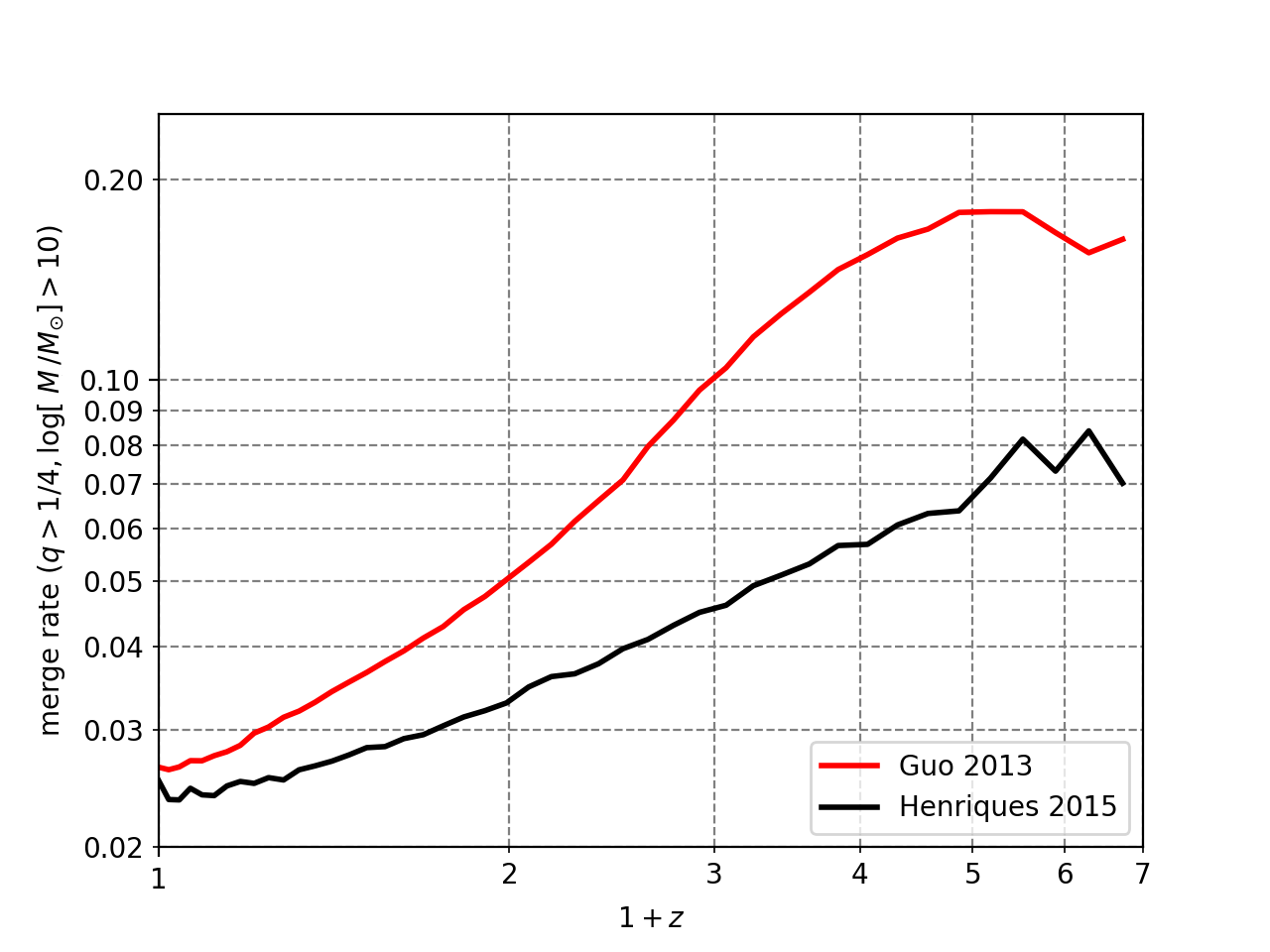}
  \caption{\small  Merger rate, defined as $n_{r}/(n_{g}\Delta t)$, for descendant galaxies with stellar mass larger than $10^{10}M_{\odot}$ and mass ratio greater than $1/4$ extracted from Guo 2013 and Henriques 2015 as a function of redshift. Here $n_r$ is the number of merge remnant galaxies, and $n_g$ is the total number of galaxies at a certain redshift. $\Delta t$ is the elapsed comvoing time between two redshift snapshots.}\label{mergerate}
\end{figure}

In Fig.~\ref{mergerate}, we plot the merger rate for galaxies as a function of redshift, for $q>1/4$ and descendant galaxies with stellar mass larger than $10^{10}M_{\odot}$ for these two models. Here $q$ refers to the stellar mass ratio between two progenitors as before. The merger rate is defined as
\beq
merger~rate=n_{r}/(n_{g}\Delta t)\;,
\eeq
where $n_r$ is the number of merger remanent at certain redshift, $n_g$ is the total number of galaxies at the same redshift, and $\Delta t$ is the comoving time step between different redshift snapshots. We can see that the merger rate in Guo 2013 is larger than that in Henriques 2015 in the whole redshift range. The enhanced factor is roughly $1.1$ around $z=0$, and reaches $2.5$ at $z=4$. In the following, we will see that the differences in the merger rate in Guo 2013 and Henriques 2015 will result in different predictions on $h_c$ according to eq. \eqref{hc}.

We briefly mention here the possible reasons for the merger rate differences between the two models. Compared to Guo 2013 and previous models, Henriques 2015 changed the time scale of reincorporation of gas ejected by supernova-driven winds, to make the galaxy massfunction and redshift dependence consistent with observations at higher redshift. The change in this process resulted in a reduction of the number and stellar mass of galaxies, and thus a smaller cross section for galaxy-galaxy collisions. This maybe the main reason for the lower merger rate of Henriques 2015. But baryonic processes considered in SAM are complex and all coupled, many parameters in Henriques 2015 has been changed, a detailed description of Henriques 2015 and Guo 2013 and their behavior in galaxy growth can be found in \cite{Henriques:2014sga}, \cite{Guo:2007wv} and \cite{Guo:2010ap}. A comparison between Henriques 2015 and De Lucia \& Blaizot 2007 \citep{DeLucia:2006szx} can be found in \cite{Vulcani:2015fka}.
The specific reasons why the merger rates of Henriques 2015 and Guo 2013 differ so significantly is still under discussion, and will be discussed in following papers.

\subsection{Results for the gravitational wave background amplitude}

\begin{figure}
  \centering
  \includegraphics[width=0.5\textwidth]{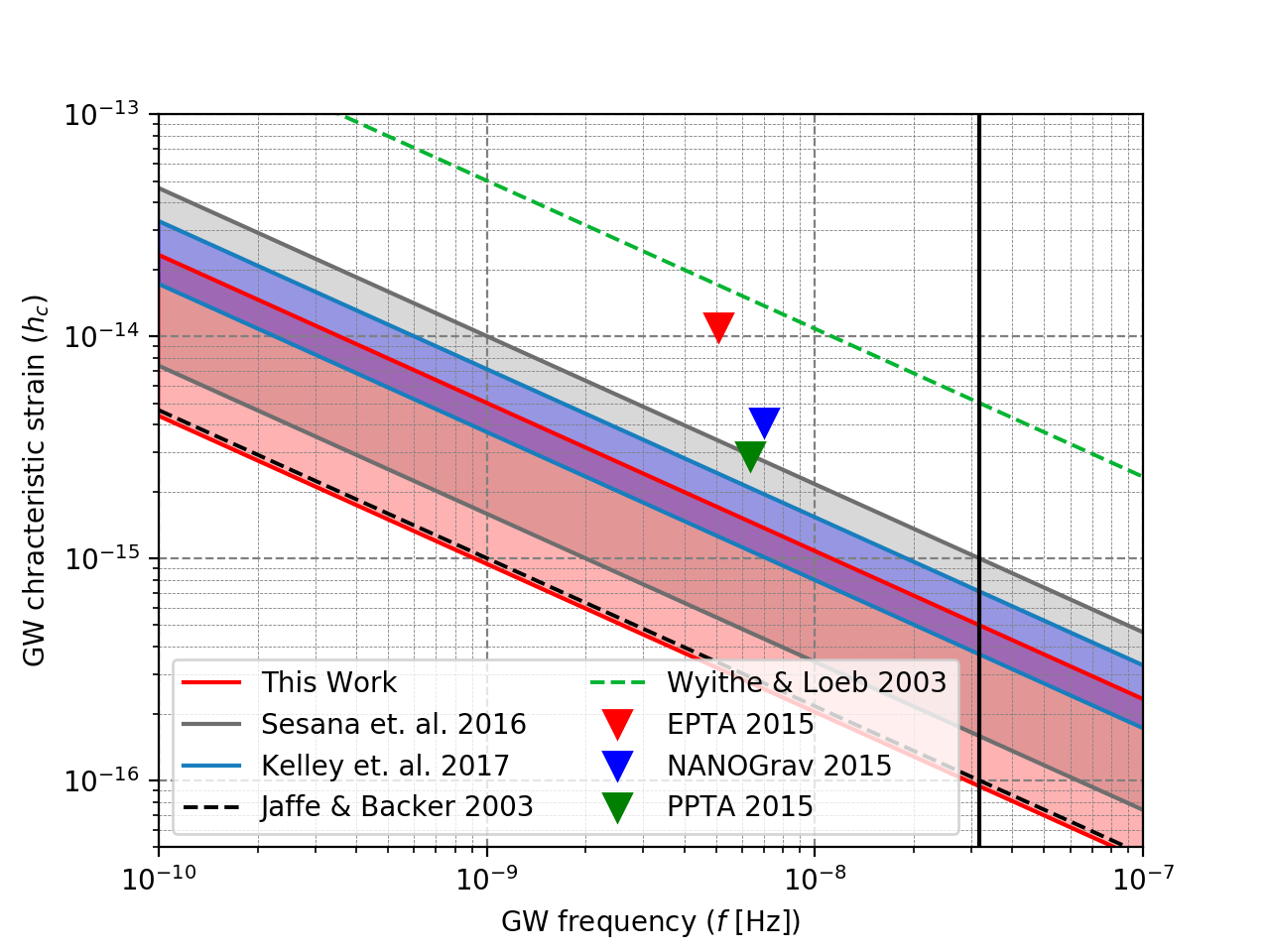}
  \caption{\small The characteristic GW strain amplitude computed in this work.
  For comparison, we also include the existed results in the literatures.
  The upper red curve is obtained from Guo 2013 model, and the lower red curve is from Henriques 2015.
  The upper limits on the GW background from PTAs are also shown. The black vertical line highlights the reference frequency $f_0=1yr^{-1}$.}\label{hcpic}
\end{figure}

After inserting the results of $d^2n/dzd\mathcal{M}$ (can be derived from the events distribution shown in Fig.~\ref{nummerger}) into eq.\eqref{hc}, we get the characteristic strain amplitude $A_{yr^{-1}}$ for the two galaxy formation models. Our results are $A_{yr^{-1}}=5.00\times10^{-16}$ (Guo 2013) and $A_{yr^{-1}}=9.42\times10^{-17}$ (Henriques 2015), respectively. In Fig.~\ref{hcpic}, we plot our results together with several predictions made by previous papers,
including the predictions derived in \cite{Jaffe:2002rt}, which used phenomenological galaxy merger rate from CNOC2 and CFGRS redshift surveys and $M_{bh}-$spheroid mass relationship;
\cite{Wyithe:2002ep}, which used semi-analytic calculation of the merger rate history at all redshifts, and phenomenological $M_{bh}-$velocity dispersion relationship;
\cite{Kelley:2016gse}, which used coevolved populations of SMBH and galaxies from hydrodynamic, cosmological simulations; and \cite{Sesana:2016yky}, which, as in \cite{Sesana:2012ak}, utilised several observed galaxy mass functions and pair counts to phenomenological SMBH-host relations, and assuming merger timescale prescriptions derived by detailed hydrodynamical simulations of galaxy mergers, but selection bias is considered in SMBH-galaxy mass relationship.
We summerize these predictions for characteristic strain amplitude in Table.~\ref{table1}, with a brief summary of the methods they used.

Several recent pulsar timing array (PTA) upper limits (summarized in Table.~\ref{table2}), such as EPTA \citep{Lentati:2015qwp}, NANOGrav \citep{Arzoumanian:2015liz}, PPTA \citep{Shannon:2015ect},  are also included for reference.
First of all, as shown in Fig. \ref{hcpic}, both our results are still below the most stringent observational upper limits.
Secondly, we see that the characteristic strain amplitude derived from Guo 2013 (upper red curve) is well consistent with most of previous results, while $h_c$ from Henriques 2015 (lower red curve) is a little bit lower than the result from \cite{Jaffe:2002rt} (black dashed curve), and is the lowest of all the predictions shown in the figure. The low prediction of GW strain amplitude in Henriques 2015 is a result of its low merger rate and low chirp mass distribution.
And more importantly, our results reveal the fact that \emph{difference in the galaxy merger rate between Guo 2013 and Henriques 2015 (shown in Fig. \ref{mergerate}), results in a factor $5$ deviation in the GW strain amplitude (shown in Fig. \ref{hcpic}).}

\begin{table*}
\centering
\begin{tabular}{ccc}
\hline
\hline
 & $\log{A_{yr^{-1}}}$ & Methods\\
\hline
\hline
This work & -15.30 & SAM, Guo 2013 \\
\hline
This work & -16.05 & SAM, Henriques 2015 \\
\hline
\cite{Kelley:2016gse} & -15.15 & Cosmo-Hydro \\
\hline
\cite{Sesana:2016yky} & $-15.4\pm0.4$ & phenomenological, bias considered \\
\hline
\cite{Sesana:2012ak} & $-15.1\pm0.3$ & phenomenological\\
\hline
\cite{Wyithe:2002ep} & -14.3 & SAM+phenomenological \\
\hline
\cite{Jaffe:2002rt} & -16 & phenomenological \\
\hline
\end{tabular}
\caption{Predictions on the GW Background, as well as the methods, from this work and literatures. Here `SAM' refers to semi-analytical models. `Cosmo-Hydro' refers to hydrodynamic simulations. `Phenomenological' refers to phenomenological formula extracted from observations.
}\label{table1}
\end{table*}

\begin{table*}
\centering
\begin{tabular}{cccc}
\hline
\hline
 PTA & $A_{yr^{-1}}$ & $A_{f_{0}}$ & $f_0[yr^{-1}]$ \\
\hline
\hline
EPTA \cite{Lentati:2015qwp} & $3.0\times10^{-15}$ & $1.1\times10^{-14}$ & 0.16 \\
\hline
NANOGrav \cite{Arzoumanian:2015liz} & $1.5\times10^{-15}$ & $4.1\times10^{-15}$ & 0.22 \\
\hline
PPTA \cite{Shannon:2015ect} & $1.0\times10^{-15}$ & $2.9\times10^{-15}$ & 0.2 \\
\hline
IPTA \cite{Verbiest:2016} & $1.5\times10^{-15}$ & $ - $ & $ - $ \\
\hline
\end{tabular}
\caption{Upper Limits on the GW Background from Pulsar Timing Arrays. The most stringent constraint on amplitude $A_{f_0}$ given at reference frequency $f_0$ are also shown.}\label{table2}
\end{table*}

\section{Black hole clustering}
\label{sec:correlation}


\begin{figure}
\centering
\includegraphics[width=0.5\textwidth]{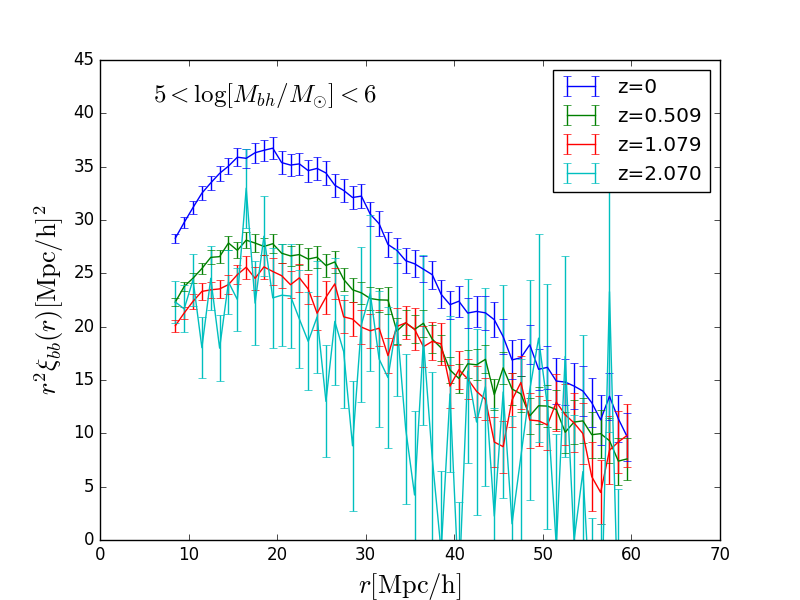}
\includegraphics[width=0.5\textwidth]{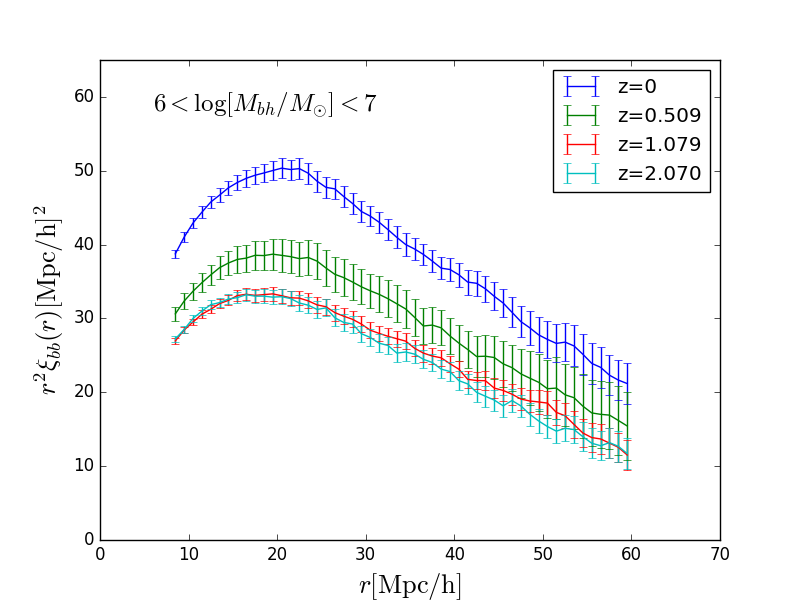}
\includegraphics[width=0.5\textwidth]{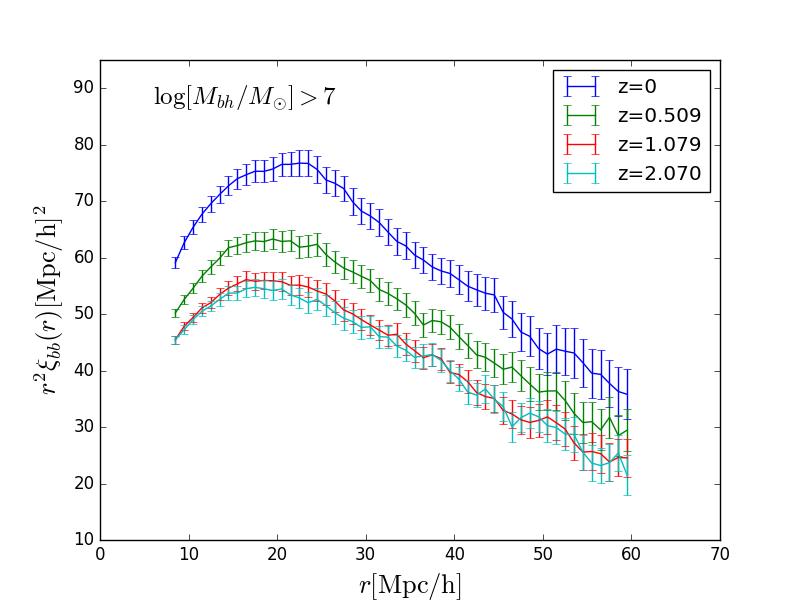}
\caption{\small The auto-2PCFs of SMBHs for different redshift-mass bins. Top panel: 2PCFs at different redshifts for $5<\log[M_{bh}/M_{\odot}]<6$. Middle panel: 2PCFs at different redshifts for $6<\log[M_{bh}/M_{\odot}]<7$. Bottom panel: 2PCFs at different redshifts for $\log[M_{bh}/M_{\odot}]>7$. Jackknife errors are also shown in the figures.}\label{correlationBH}
\end{figure}
\begin{figure}
\centering
\includegraphics[width=0.5\textwidth]{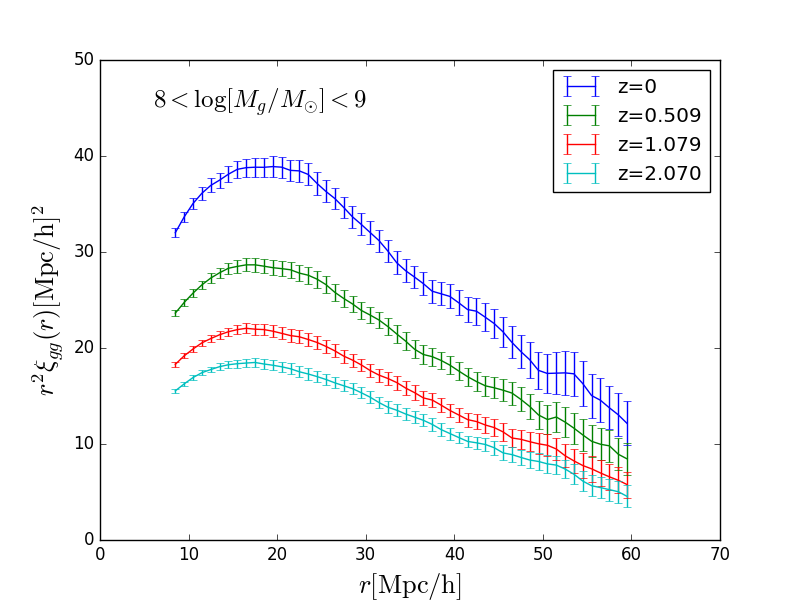}
\includegraphics[width=0.5\textwidth]{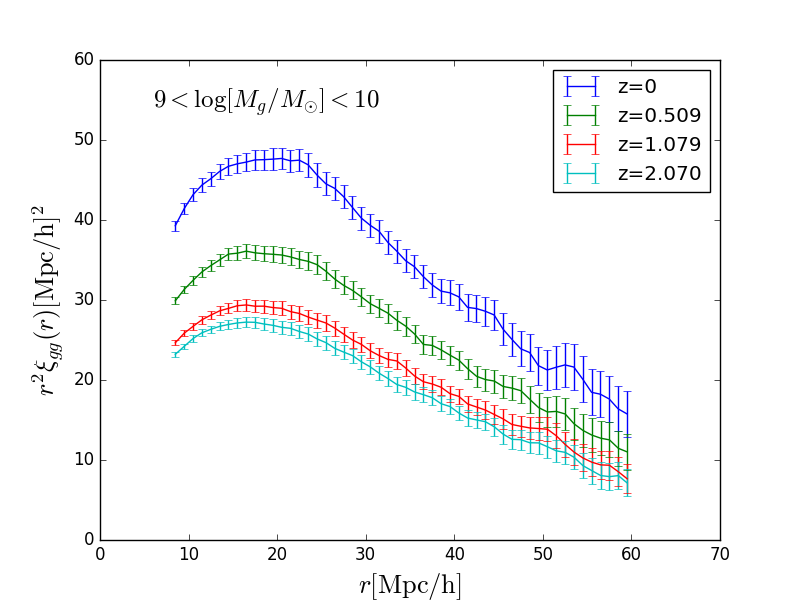}
\includegraphics[width=0.5\textwidth]{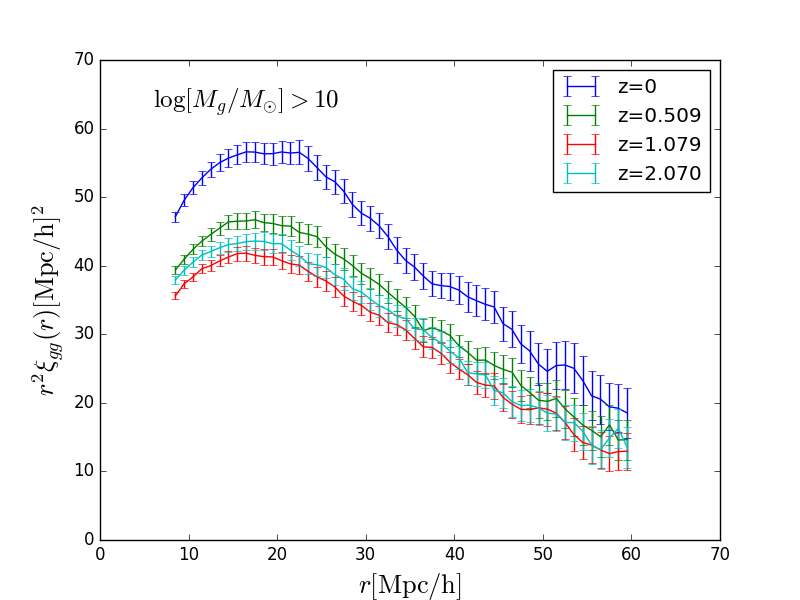}
\caption{\small The auto-2PCFs of galaxies for different redshift-mass bins. Top panel: 2PCFs at different redshifts for $8<\log[M_{g}/M_{\odot}]<9$. Middle panel: 2PCFs at different redshifts for $9<\log[M_{g}/M_{\odot}]<10$. Bottom panel: 2PCFs at different redshifts for $\log[M_{g}/M_{\odot}]>10$.}\label{correlationG}
\end{figure}


\begin{figure}
\centering
\includegraphics[width=0.5\textwidth]{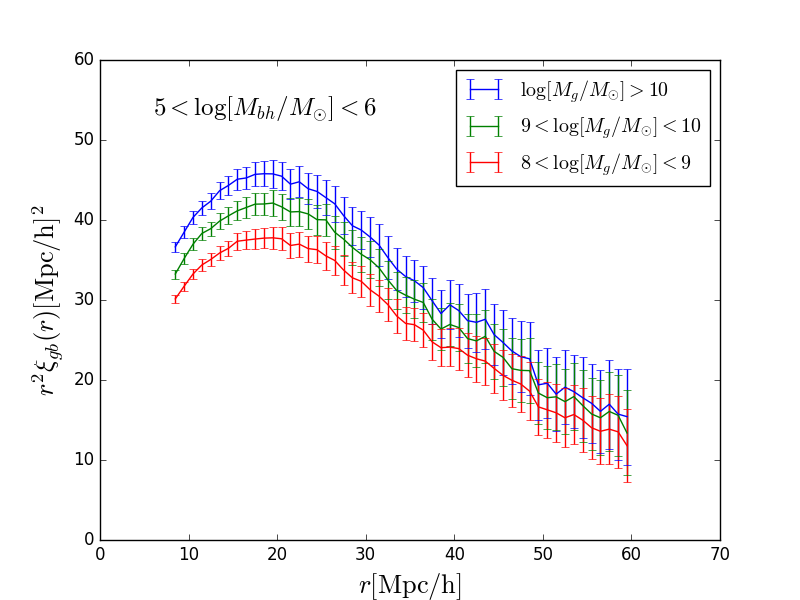}
\includegraphics[width=0.5\textwidth]{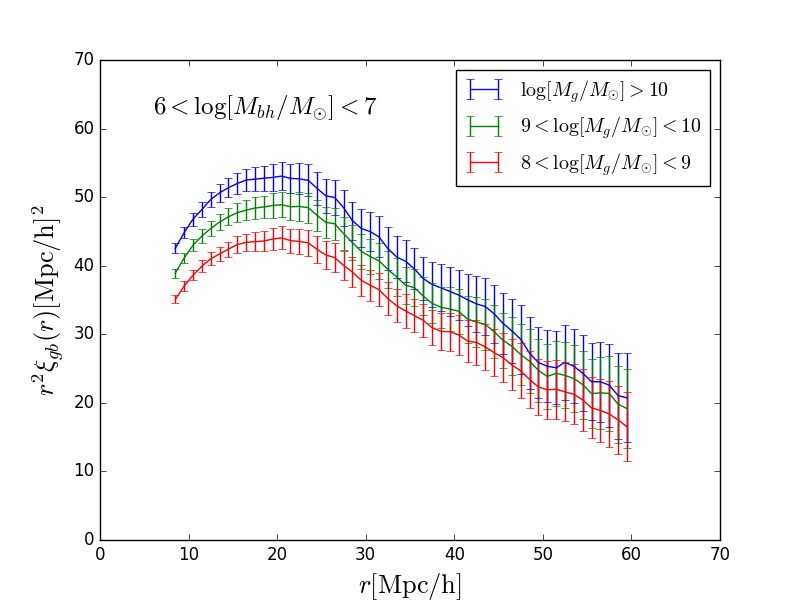}
\includegraphics[width=0.5\textwidth]{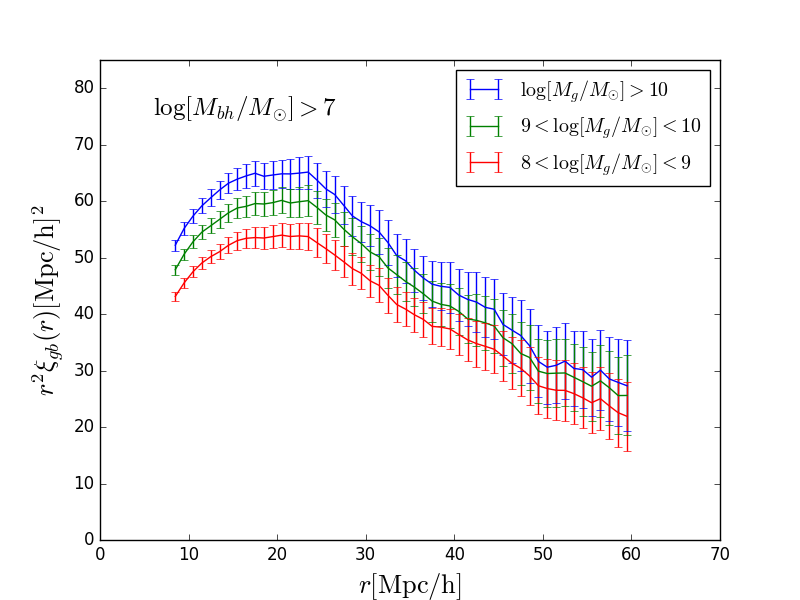}
\caption{\small The cross-2PCFs between SMBHs and galaxies at $z=0$. Top panel: the cross-2PCFs of different galaxy mass bins with $5<\log[M_{bh}/M_{\odot}]<6$. Middle panel: the cross-2PCFs of different galaxy mass bins with $6<\log[M_{bh}/M_{\odot}]<7$. Bottom panel: the cross-2PCFs of different galaxy mass bins with $\log[M_{bh}/M_{\odot}]>7$. }\label{crosscorrelation}
\end{figure}

It is generally believed that gas accretion onto SMBHs in the galaxy center is the energy source of AGN. In order to understand the relationship between growth of SMBHs and their surrounding environment, it is important to investigate the clustering
properties of both SMBHs and their host galaxies.

The auto-correlation function of AGNs was studied by using the SDSS sample \citep{Shen:2006ti,Shen:2008ez,Ross:2009sn}. They found that while no significant evolution of SDSS quasar clustering amplitude
can be seen for $z<2.5$, clustering strength does increase at higher redshift. \cite{Shen:2008ez} further studied the dependence of the two-point auto-correlation function of quasars on luminosity, BH mass, colour, and radio loudness, and found positive dependence on radio-loudness, weak or no dependence on virial BH mass for $z<2.5$. The clustering property of intermediate redshift quasars using the final SDSS III-BOSS sample was investigated in \cite{Eftekharzadeh:2015ywa}. The clustering of galaxies around AGNs in the areas of deep surveys was also investigated by \cite{Coil:2009bi} and \cite{Mountrichas:2008jf}.

Recently, cross-correlation between AGNs and galaxies was studied using large samples \citep{Donoso:2009wd,Krumpe:2011ra}. The results show that that radio-loud AGNs are clustered more strongly than radio-quiet ones, while no significant difference was found between X-ray selected and optically selected broad-line AGNs.
 \cite{Donoso:2009wd} found a positive dependence of the cross-correlation amplitude on stellar mass
$M_*$, but within a narrow range of stellar mass ($10^{11}M_{\odot}<M_*<10^{12}M_{\odot}$). The mass dependence of cross-correlation between AGN and galaxies is further investigated in \cite{Komiya:2013vja} with $9394$ AGNs for $z = 0.1-1$ over a wide BH mass ($10^{6}M_{\odot}<M_*<10^{10}M_{\odot}$). There is an indication of an increasing trend of correlation with BH mass for $M_{bh}>10^8M_{\odot}$, while no BH mass dependence for $M_{bh}\lesssim10^8M_{\odot}$. \cite{Shirasaki:2015lu} studied the cross-correlation with updated UKIDSS catalog and reconfirmed the findings of \cite{Komiya:2013vja} that the clustering of galaxies around AGNs with the most massive SMBH is larger than those with less massive
SMBH, and AGN bias was derived for each BH mass group.

On the simulation side, the clustering property of quasars or SMBH are also investigated in several works. \cite{Oogi:2015cxi} studied the clustering properties of quasars via quasar bias. \cite{DeGraf:2016bbt} also studied the clustering properties of SMBHs for Illustris simulation.


In order to justify/falsify the galaxy formation model, here we investigate the clustering properties of both SMBHs and galaxies, and corresponding cross-correlations between them.
To do this, we use the mock catalog produced by Guo 2013, which allows us to study their dependence over a wide range of redshift and SMBH/galaxy mass with large samples.
We selected four redshift snapshots which contains a total of 8668809 SMBHs and 51538704 galaxies lying in a box size of $[500\mathrm{Mpc}/h]^3$ created by applying the Guo 2013 SAM model to the Millennium simulation.
The Millennium simulation was created in a cosmology of $(\Omega_b, \Omega_m, \sigma_8, h) = (0.0456, 0.273, 0.809, 0.704)$ using $2160^3$ particles.
For the data completeness, the redshift and SMBH/galaxy mass range we are interested in are $0<z<2.07$, $M_{bh}>10^6M_{\odot}$ and $M_{g}>10^8M_{\odot}$, where $M_{bh}$ is the mass of SMBHs, and $M_g$ is the stellar mass of galaxies.
We abandoned the data in the lower mass and higher redshift regime because of the limited resolution and small catalog size, respectively.

The galaxy and SMBH clustering is adequately described by the spatially isotropic two-point correlation function (2PCF), which is computed by the excess of data-data number counts relative to those of random pairs \citep{Landy:1993yu}
\beq\label{correlation}
\xi(r)=\frac{DD(r)-2DR(r)+RR(r)}{RR(r)}\;,
\eeq
where $DD(r)$ is the data-data number counts at different clustering scales, and $DR(r)$, $RR(r)$ being the data-random, random-random number counts, respectively.

To calculate the errors on 2PCFs, we adopt the `delete one jackknife' method. We divide the full galaxy/SMBH sample into 125 sub-boxes, and calculate the 2PCF for $125$ sub-samples consisting all but the $k$-th sub-box. This allows us to construct a jackknife defined covariance matrix
\beq
C(r_i,r_j)=\frac{124}{125}\sum_{k=1}^{125}[\overline{\xi}(r_i)-\xi_k(r_i)][\overline{\xi}(r_j)-\xi_k(r_j)]\;,
\eeq
where $\xi_k(r)$ refers to the value of the correlation obtained by omitting the $k^{\mathrm{th}}$ sub-box, $\overline{\xi}(r_i)$ is the average correlation value for all the subsamples, and $i$, $j$ refers to the $i^{\mathrm{th}}$ and $j^{\mathrm{th}}$ bins, respectively. The binned errors $\sigma_i$ can be obtained from the diagonal elements of the covariance metric as
\beq
\sigma^2_i=C_{i,i}\;.
\eeq

To perform the calculation of correlation, we divide the SMBH and galaxy simulation sample into $12$ redshift-mass groups. For redshift, we choose $4$ snapshots (shown in eq. \ref{eq:zbin}). SMBH and galaxy masses are divided into $3$ bins, shown in eq. (\ref{eq:bhmassbin}) and (\ref{eq:gmassbin}), respectively.
\beq
z=0\;, 0.509\;, 1.079\;, 2.070\;.\label{eq:zbin}
\eeq

\beq
5<\log[M_{bh}/M_{\odot}]<6\;, 6<\log[M_{bh}/M_{\odot}]<7\;, \log[M_{bh}/M_{\odot}]>7\;.\nonumber\\
\label{eq:bhmassbin}
\eeq

\beq
8<\log[M_{g}/M_{\odot}]<9\;, 9<\log[M_{g}/M_{\odot}]<10\;, \log[M_{g}/M_{\odot}]>10\;.\nonumber\\
\label{eq:gmassbin}
\eeq

The auto-2PCFs are calculated in all redshift bins both for SMBHs and galaxies. The corresponding results are shown in Fig.~\ref{correlationBH} (SMBH) and Fig.~\ref{correlationG} (galaxy). The $z=2.07$ 2PCF in the first panel of Fig.~\ref{correlationBH} has larger error bars due to the limited SMBH samplings in the corresponding mass and redshift range.
Comparing the 2PCFs in different plots with the same redshift and same population (SMBHs or galaxies), we see that 2PCFs have a positive dependence on mass for both SMBHs and galaxies, namely, SMBHs and galaxies are more correlated if they are more massive. This is consistent with the structure formation scenario, which states that more massive objects are formed in higher dense regions and thus have stronger clustering strength.

Another behavior is that the 2PCFs seems to increase faster with mass at higher redshifts than at lower redshifts for both SMBHs and galaxies. In particular, in the most massive cases, \emph{i.e.} the last plot of Fig.~\ref{correlationG}, we see that the correlation of the highest redshift bin has run over that of the lower redshift ones. The reason is that, at higher redshift galaxies/SMBHs as a whole population are less massive, by imposing a same mass cut at all redshifts, at high redshifts we are systematically selecting more biased objects, who reside in larger density contrast regions and thus have larger clustering strength.

In most panels of Fig.~\ref{correlationBH} and Fig.~\ref{correlationG}, we see a clear redshift dependence of the 2PCF amplitude, namely, the clustering of both SMBHs and galaxies is enhanced at lower redshifts. This is due to the gravitational growth of structure with the evolution of time.
Interestingly, this behavior is not reported in previous papers with SDSS samples \citep{Shen:2006ti,Shen:2008ez,Ross:2009sn}, where they didn't find significant evolution or just a weak positive redshift dependence of clustering for $z<2.5$ .
This should result from the difference between the properties of the two samples.
The SDSS sample analyzed in \cite{Ross:2009sn} is a flux limited sample,
so the bias of objects rapidly increases at higher redshifts (where only very bright objects can be observed);
the sparseness of the sample also weakens the power of statistics and makes the detection of redshift evolution relatively difficult.
The samples considered in this analysis are denser and have constant mass cut at all redshifts.
As a result, we can clearly detect the redshift evolution of clustering strength.
In all, the combination of structure growth and selection effect determines how the amplitude of correlation evolves with redshift.

As for the shape of correlation functions, they are not evolving a lot at $z>1$, but starts to have larger slope at lower redshifts. This is expectable, since the non-linear growth of structure at later epoch results in enhancement of clustering in relatively small scales. Due to the same reason, correlation functions of more massive objects have larger slope than less massive objects results (more massive objects are more biased and they experienced more non-linear growth of structure). Comparing the 2PCFs between SMBHs and galaxies, since SMBHs reside in galaxies, the apparent clustering of SMBHs is actually a result of galaxy clustering, so it is not surprising that the amplitude and shape of SMBH correlation functions are similar to those of galaxies. This analysis shows that, very roughly, SMBHs and galaxies, with galaxy mass $10^2\sim10^3$ larger than SMBH mass, have similar pattern of clustering (strength and scale dependence).

The results for cross-correlation between different SMBH and galaxy mass bins at $z=0$ are shown in Fig.~\ref{crosscorrelation}. We calculate the cross-correlation between each of the three mass bins of SMBH and galaxy, so there are nine 2PCFs in total. And each panel in Fig.~\ref{crosscorrelation} is for a fixed SMBH mass. We see in each plot, the cross-correlations become larger for higher galaxy stellar mass. Comparing the 2PCFs in different plots with the same $M_g$, we also find that the correlation amplitude become larger for higher SMBH mass. That is, in summary, the cross-correlation between SMBH and galaxies has a positive dependence on both SMBH and galaxy mass, which is consistent with previous results from observed AGN and galaxy samples \citep{Donoso:2009wd,Komiya:2013vja,Shirasaki:2015lu}.

\section{Anisotropy of the gravitational wave background}
\label{sec:anisotropy}
The spatial inhomogeneities of the SMBH clustering should produce some amount of anisotropies in the GW background signal.
In this section we will investigate the anisotropic property of the GW background generated by binary SMBH sources with a mock merger event catalog produced by Guo 2013. Anisotropy can be investigated by decomposing the energy density of the GW background \citep{Taylor:2013esa, Kuroyanagi:2016ugi}, which now is a function of the angular coordinate $\hat\Omega$ on the 2D sphere, in terms of the spherical harmonic functions as
\beq
\rho(\hat\Omega)=\sum^{\infty}_{l=0}\sum^{l}_{m=-l}c_{lm}Y_{lm}(\hat\Omega)\;.
\label{rhodecompose}
\eeq
So the spherical harmonic coefficients $c_{lm}$ can be calculated by integrating the GW background signal over all directions
\beq
c_{lm}=\int d\hat\Omega \rho(\hat\Omega)Y_{lm}(\hat\Omega)\;.
\label{clm}
\eeq
The statistical isotropic angular power spectrum of GW background density reads
\beq
C_l=\sum_m|c_{lm}|^2/(2l+1)\;.
\label{spectrum}
\eeq
In the realistic case for the GW background generated by SMBH binaries, the number of sources is always discrete and finite, so eq.~\eqref{clm} can be replaced by a discrete form:
\beq
c_{lm}=\sum^N_{i=1}\rho_iY_{lm}(\hat\Omega_i)
\label{clmdiscrete}
\eeq
where $\rho_i$ is the energy density generated by the $i$th source, and is given by
\beq
\rho=\frac{8}{5\pi D_L(z)^2}[\pi\mathcal{M}f(1+z)]^{10/3}\;.
\label{rho}
\eeq

\begin{figure}
\centering
\includegraphics[width=0.5\textwidth]{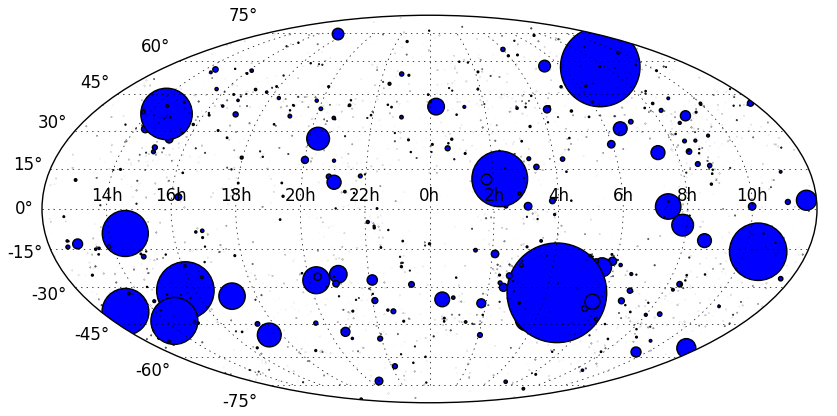}
\caption{\small The skymap of all the SMBH binary sources from the mock merger event catalog of Guo 2013. 8426 SMBH binaries is contained in the catalog in total. Frequency is assigned to each binary according to $P(\mathrm{ln}f)\propto\mathcal{M}^{-5/3}f_r^{-8/3}$, and the relative size of each source is a indication of the GW energy density.}\label{skymap}
\end{figure}

\begin{figure}
\centering
\includegraphics[width=0.5\textwidth]{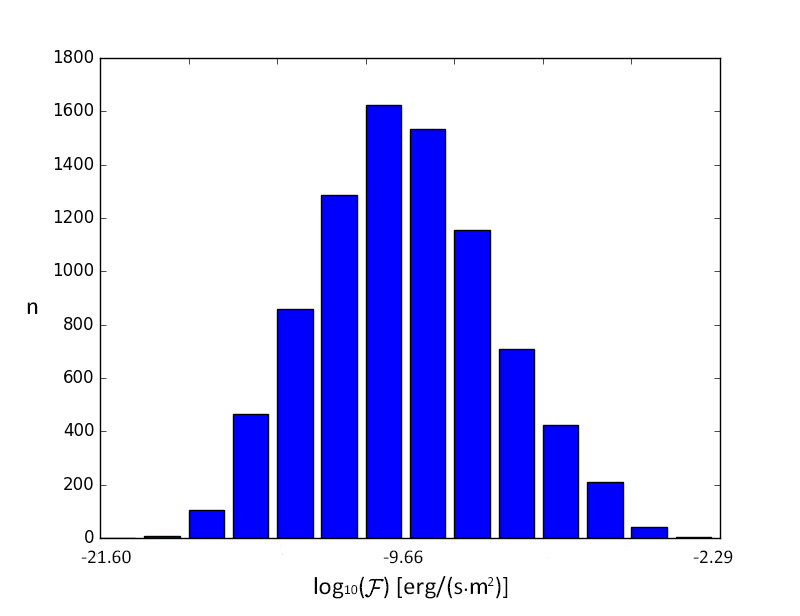}
\caption{\small The energy flux distribution histogram for all the GW sources from the mock merger event catalog of Guo 2013. Here $\mathcal{F}$ is the energy flux of the GWs from each source, and $n$ is the number of sources that fall into a corresponding logarithm flux bin.}\label{histogram}
\end{figure}

Our mock catalog from Guo 2013 contains 8426 galaxy merger events with binary SMBHs in total, each merger event carries information about their location and the mass of the progenitor SMBHs. To calculate the energy density from eq.~\eqref{rho}, we still need to assign frequency to each of the SMBH binaries inside the merging galaxies. We assume that the seperation between the two black holes of the SMBH binaries lies between $10^{-2}\mathrm{pc}$ and their inner most stable orbit, which corresponds to minimum and maximum frequency, respectively. We assign frequency to each SMBH binary according to a normalized probability that is proportional to eq.~\eqref{dtdf}, i.e.,
\beq\label{probability}
P(\mathrm{ln}f)=C\times\frac{5}{64\pi^{8/3}}\mathcal{M}^{-5/3}f_r^{-8/3}\;,
\eeq
where C is a normalization constant. This assignment has a meaning that the probability for a logarithmic frequency bin, that a SMBH binary may lie in, is proportional to the evolution time the binary spends in that frequency bin. Now with frequency, progenitor mass and location, we can calculate energy density for each SMBH binary. We plot a skymap of all the SMBH binary sources from the mock catalog in Fig.~\ref{skymap}, the relative size of each source is a indication of the GW energy density. We also plot an energy flux distribution histogram for all the sources in Fig.~\ref{histogram}, where the GW energy flux $\mathcal{F}$ is related to the energy density by $\mathcal{F}=c\rho$, and $c$ is the speed of light. As can be seen from the skymap and the histogram, there're several bright sources shine upon the majority dim ones, and may give a major contribution to the result of the angular power spectrum.

\begin{figure}
\centering
\includegraphics[width=0.5\textwidth]{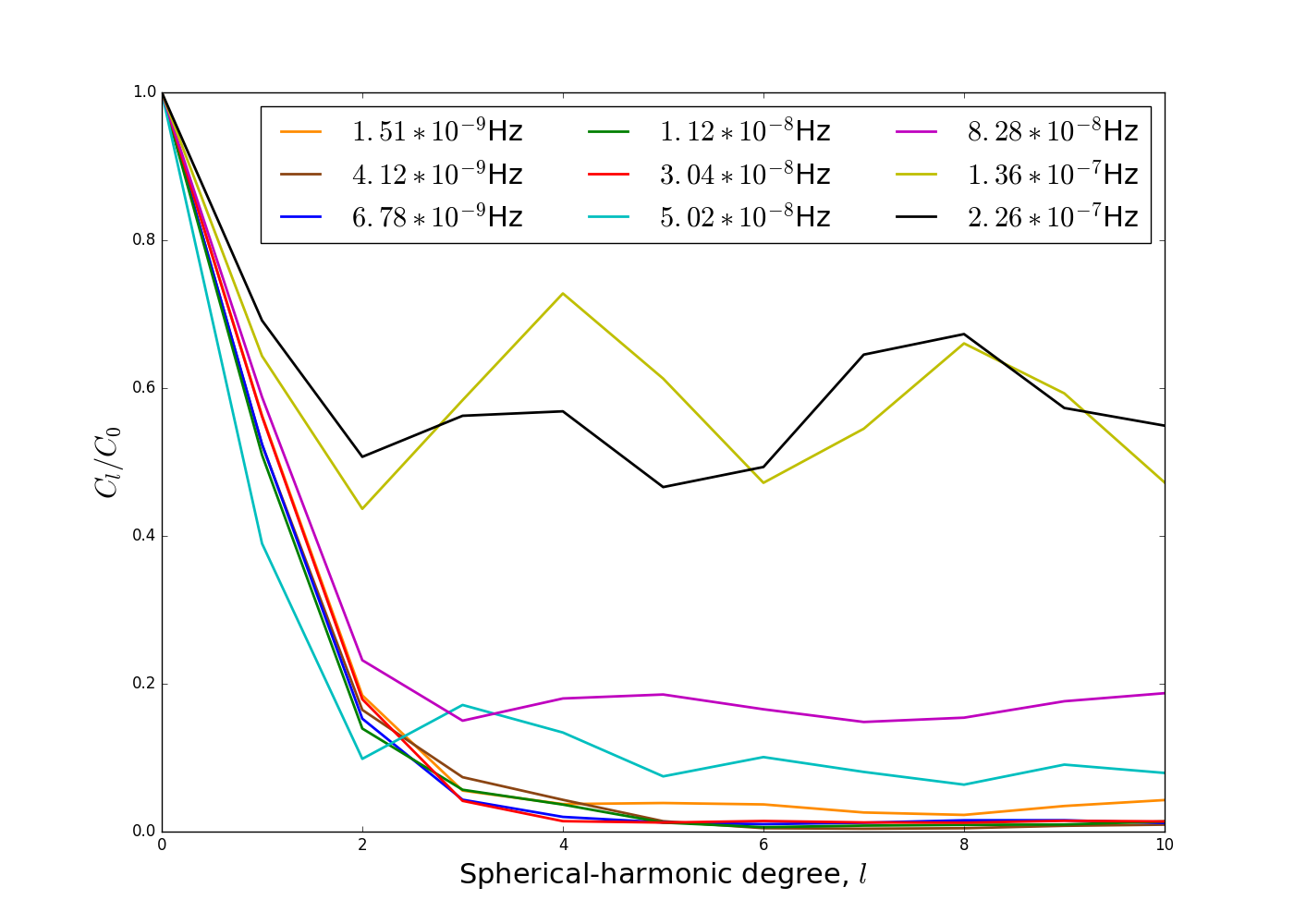}
\caption{\small Our result for the anisotropic power spectrum normalized by the monopole component, i.e., $C_l/C_0$ up to $l=10$. The results are sorted into different frequency bins from $f=1.51\times 10^{-9}\mathrm{Hz}$ to $f=2.26\times 10^{-7}\mathrm{Hz}$.}\label{powerspectrum}
\end{figure}

Now, we can calculate the angular power spectrum for the SMBH binaries that fall into the PTA band. We calculate the first ten multiples of the angular power spectrum for each frequency bin, and our result is shown in Fig.~\ref{powerspectrum}. From the plot we see that, in the whole frequency range, $C_l$ can have about 10\% to 60\% the power of the monopole component. This is a considerable level of anisotropy, and may come from the several bright sources which can be seen from Fig.~\ref{skymap}. For $l=1$, $C_1$ is in the range of 40\% to 70\% the amount of the monopole component. For higher multiples, $C_l$ drop quickly for in low frequency range $f<8.28\times10^{-8}\mathrm{Hz}$. Their amplitudes are of order $\mathcal{O}(10\%)$ w.r.t. the power of the monopole component.
As of the high frequency modes, such as $f=1.36\times10^{-7}\mathrm{Hz}$ and $f=2.26\times10^{-7}\mathrm{Hz}$, the higher multiples are of order $\mathcal{O}(60\%)$.
This is because the number of low frequency SMBH binaries are much more than those in high frequency band. Hence, the spatial distribution anisotropy is obviously larger for the small population.
For example, in frequency bin above $2.26\times10^{-7}\mathrm{Hz}$, there's only $18$ merger events.

\section{Conclusions}
\label{sec:conclusion}
In this paper, we investigated the co-evolution of supermassive black holes (SMBHs) with galaxies by studying the stochastic gravitational wave background radiation generated by SMBH merger and the SMBH/galaxy clustering, namely, the two point auto- and cross-correlation functions, by using the mock catalogs generated by the semi-analytic model (SAM) of galaxy formation. For SAM, we utilize the Munich model, which is based on the sub-halo merger trees built from the Millennium/Millennium-II simulations, and applied to WMAP (Guo 2013), Planck cosmologies (Henriques 2015) .

For the stochastic gravitational wave background, we firstly compared the mass-redshift distribution of SMBH merger events and galaxy merger rate for Guo 2013 and Henriques 2015 models.
We found that SMBHs merger is less massive and less efficient in Henriques 2015 than in Guo 2013 and galaxy merger rate for Guo 2013 is higher than that for Henriques 2015 in the whole simulated redshift range.
Quantitatively, the maximum of differential event number of SMBH merger locates around $z=0.5$, $\log\left(\mathcal{M}/M_{\odot}\right)=7$ (Guo 2013) and $\log\left(\mathcal{M}/M_{\odot}\right)=5.5$ (Henriques 2015), respectively.
And it reaches above $2100$ in Guo 2013, while it is below 2000 in Henriques 2015. As for the galaxy merger rate, with stellar mass ratio $q>1/4$ and stellar mass greater than $10^{10}M_{\odot}$, Guo 2013 model is systematically higher than Henriques 2015. The enhanced factor is roughly $1.1$ around $z=0$, and reaches $2.5$ at $z=4$.

We then predicted the characteristic strain amplitude of GW background for Guo 2013 and Henriques 2015 model to be $A_{yr^{-1}}=5.00\times10^{-16}$ and $A_{yr^{-1}}=9.42\times10^{-17}$, respectively.
We shall emphasize that, the GW amplitude is very sensitive to the galaxy merger rate. The difference in galaxy merger rate between Guo 2013 and Henriques 2015 (shown in Fig. \ref{mergerate}), results in a factor $5$ deviation in the GW strain amplitude (shown in Fig. \ref{hcpic}). Furthermore, we compared our result with those in literatures with different methods. We found that $h_c$ from Guo 2013 is more closer to other studies while
Henriques 2015 model gives the lowest prediction on the GW signal.

We now briefly discuss the simplifications we have used. In calculating the GW background amplitude, we have neglected environmental effects. We assumed that the orbits of the binary SMBHs are all circular and they emit GWs through quadrupole formula. Furthermore, we have assumed that binary SMBHs merge only through GW emission and with 100\% efficiency. In the reality cases, merger processes are more complex than the situations we have considered. First of all, in realistic situations, the interactions between binary SMBHs and gas or stellar environment may increase the binaries' eccentricity and the binary SMBHs emitting GWs in the PTA band may have non-negligible eccentric orbits \citep{Armitage:2005xq,Matsubayashi:2005eg,Berentzen:2008yw,Sesana:2010qb,Preto:2011gu,Khan:2011gi}, and the GW background spectrum will be changed and deviate from the simple $-2/3$ case here \citep{Enoki:2007sl,Ravi:2014aha,Sesana:2014wta}. On the other hand, binary SMBHs inside galaxies generally undergo dynamical friction, loss-cone stellar scattering, viscous drag to reach the GW domain regime. Among these processes, the efficiency of loss-cone scattering is still quite uncertain, as is pointed out by \cite{Merritt:2013vk}, and generated the famous ``last parsec problem" \citep{Milosavljevic:2002ht,Merritt:2004gc}. We have assumed binary SMBHs coalesce simultaneously when host galaxies merge on kpc scale, for simplicity. Although our treatment is simpler than some of the studies in the literature, but our approach is fast and flexible, hence, we can update our result once the more reliable SAM model are presented.

For clusterings, we calculated the spatially isotropic two point auto- and cross-correlation functions (2PCFs) for both SMBHs and galaxies by using the mock catalogs generated from Guo 2013 model.
We studied their dependence through a wide range of redshift as well as black hole and galaxy stellar mass.
We showed that all 2PCFs have positive dependence on both SMBH and galaxy mass.
And there exist a significant time evolution in both the SMBH and galaxy 2PCFs due to the gravitational growth of structure with the evolution of time, namely, the clustering effect is enhanced at lower redshift. Interestingly, this behavior is not reported in the previous AGN samples from SDSS survey, which should result from the increase of bias objects at higher redshifts and the sparseness of their sample.
As for the shape of the 2PCFs, we found they always have larger slope at lower redshifts due to non-linear growth of structure and enhancement of clustering in relatively small scales at later epoch. We also showed that roughly, SMBHs and galaxies, with galaxy mass $10^2\sim 10^3$ larger than SMBH mass, have similar pattern of clustering.

For the GW background anisotropy, we calculated the angular power spectrum up to $l=10$ with a mock merger event catalog form Guo 2013. The catalog contains $8426$ SMBH binaries in total. We assign frequency to the SMBH binaries by assuming a probability proportional to their evolution time in the corresponding logarithmic frequency bin. We found a considerable amount of anisotropy. The corresponding angular power spectrum of the first ten multiples are about $10\%$ to $60\%$ w.r.t. the monopole component in the whole frequency range of $1.51\times10^{-9}\mathrm{Hz}$ to $2.26\times10^{-7}\mathrm{Hz}$. Several bright sources of the catalog offered the major contribution to this level of anisotropy.

Environmental effects may also affect the results of anisotropy. Recall that we have used a probability proportional to eq.~\eqref{dtdf}, which is a purely quadrupole formula. According to this probability, a binary has much a greater chance to be found in the low frequency regime than in the high one. Environmental effects, if taken into consideration, would accelerate the merging process of the two SMBHs before the GW dominated era, and suppress the evolution time difference between frequency bands. Thus the opportunity to have a binary in the high frequency regime shall increase. The corresponding result for the angular power spectrum with high frequency should be suppressed due to a larger binary population.

In this paper, we studied the GW background signal (both isotropic and anisotropic components) and clustering status of SMBHs/galaxies, respectively. Since the clusterings of SMBHs and GW radiation are both the results of galaxy formations, these two aspects are tightly related with each other. Our paper is a tentative trial where these two aspects are investigated. In the future, we will consider the possible relationship between the GW background anisotropic signal and the galaxy/SMBH clustering status, or the use of the anisotropic component to distinguish the different scenario of production of GW background. We leave these for future studies.

As the PTA experiments are constantly improving their sensitivity, the first detection of background GW signal may not be far in the future. The recent large sky surveys are also aiming at broader dynamical ranges and larger sample sizes. All these experimental improvements are providing us new ways to study the SMBH growth and its co-evolution with galaxy. It is hopefully that our results may offer clue to the theoretical progress in relative fields. On the other hand, the comparison of our results with observations may help improve the galaxy formation model building.

\section*{Acknowledgements}
We thank Qi Guo for useful discussion.
QY and BH are supported by the Beijing Normal University Grant under the reference No. 312232102 and by the National Natural Science Foundation of China Grants No. 210100088.
BH is also partially supported by the Chinese National Youth Thousand Talents Program under the reference No. 110532102 and the Fundamental Research Funds for the Central Universities under the reference No. 310421107.


\bibliographystyle{mnras}
\bibliography{mnras_SMB}



\bsp	
\label{lastpage}
\end{document}